\documentclass[
]{ceurart}
\pdfoutput=1
\usepackage{microtype}

\begin{document}
\begin{sloppypar}

\copyrightyear{2024}
\copyrightclause{Copyright for this paper by its authors.
  Use permitted under Creative Commons License Attribution 4.0
  International (CC BY 4.0).}

\conference{IR4U2@ECIR’24: 1st workshop on Information Retrieval for Understudied Users, Glasgow, Scotland}

\title{Inclusive Design Insights from a Preliminary 
 Image-Based Conversational Search Systems Evaluation}

%
\author[1]{Yue Zheng}[%
orcid=0009-0000-8095-9569,
email=yz15u22@soton.ac.uk,
]

\author[2]{Lei Yu}[%
email=ly3u21@soton.ac.uk,
]

\author[3]{Junmian Chen}[%
email=jc13u22@soton.ac.uk,
]

\author[4]{Tianyu Xia}[%
email=tx2n20@soton.ac.uk,
]

\author[5]{Yuanyuan Yin}[%
email=y.yin@soton.ac.uk,
]

\author[6]{Shan Wang}[%
email=shan.wang@soton.ac.uk,
]

\author[7]{Haiming Liu}[%
orcid=0000-0002-0309-3657,
email=h.liu@soton.ac.uk,
]
\address{University of Southampton, Southampton, UK}

\begin{abstract}
  The digital realm has witnessed the rise of various search modalities, among which the Image-Based Conversational Search System stands out. This research delves into the design, implementation, and evaluation of this specific system, juxtaposing it against its text-based and mixed counterparts. A diverse participant cohort ensures a broad evaluation spectrum. Advanced tools facilitate emotion analysis, capturing user sentiments during interactions, while structured feedback sessions offer qualitative insights. Results indicate that while the text-based system minimizes user confusion, the image-based system presents challenges in direct information interpretation. However, the mixed system achieves the highest engagement, suggesting an optimal blend of visual and textual information. Notably, the potential of these systems, especially the image-based modality, to assist individuals with intellectual disabilities is highlighted. The study concludes that the Image-Based Conversational Search System, though challenging in some aspects, holds promise, especially when integrated into a mixed system, offering both clarity and engagement.
\end{abstract}

\maketitle
\section{Introduction}

The rapid development of conversational search engines has completely changed the field of information retrieval. A new era of natural language-based tailored search experiences is being ushered in by platforms like ChatGPT\cite{ray2023chatgpt}. However, users with disabilities or cognitive impairments face challenges when using standard conversational search engines, which frequently require users to possess extensive linguistic talents and awareness of the engine's intricacies. These people may have linguistic restrictions that make it more difficult for them to be receptive and expressive. The situation is further complicated by the paucity of data pertaining to the various challenges linked to cognitive impairments.

Currently, available tools to assist these users encompass assistive devices, text-to-speech technologies for digital content narration, and basic messaging systems integrating voice, gestures, or sign language\cite{light2019new}\cite{liu2010applying}. In this paper, we develop text-based, image-based, and mixed search engines with various tasks assigned to 21 participants on these platforms. By integrating sensor technologies to capture physiological signals, the aim becomes clear: determining the optimal search strategy for particular user groups to foster a more inclusive and accessible search interface.

The key component of this strategy is the conversion of textual results from conversational search engines into visual formats, which facilitates information understanding. Simultaneously, sensors are utilized to naturally record user input, removing the requirement for human intervention. The objective of this paradigm shift is to increase the accessibility of information, particularly for individuals with disabilities, thereby improving their independence and expanding their knowledge.

This research centers on an image-based conversational search system that leverages sensor data, like gestures and eye movements, to gauge user satisfaction. This feedback refines subsequent searches, fostering a more adaptive and user-centric experience.

\section{Related Work}
Recent research has delved deeply into the evolution and utility of voice assistants and conversational interfaces\cite{kaushik2022exploratory}. The emergence and significance of voice assistants like Alexa, Siri, and Cortana, emphasizing their integration in various sectors\cite{hoy2018alexa}. Additionally, spoken conversational search sheds light on user interactions in speech-only search tasks\cite{trippas2017people}. The research from both 2017 and 2018 provides a comprehensive overview of how people interact with these systems and the challenges therein \cite{trippas2018informing}.

The usability and effectiveness of these conversational interfaces have been a pivotal area of research. Ghosh assessed the System Usability Scale's utility for evaluating voice-based user interfaces, emphasizing the nuances of voice interactions \cite{ghosh2018assessing}. In a similar vein, Avula and colleagues explored the dynamics of user engagement with chatbots during collaborative searches\cite{avula2018searchbots}. Their study presented key findings on how chatbots can be leveraged to enhance the search experience.

Conversational search systems have been studied not just for individual use but also for collaborative scenarios. Avula focused on embedding search into conversational platforms to support collaborative search, a novel approach to enhance user interaction and data retrieval\cite{avula2019embedding}. Moreover, Kiesel and his team delved into voice query clarification, underscoring the need for refining voice-based search queries to improve search outcomes\cite{kiesel2018toward}.

\section{Experiment Setup}
\subsection{System Design}
To optimize response speed and quality, the selection was made to incorporate several well-established APIs. The choice of these specific four APIs was influenced by a combination of factors: the precision of their image analysis algorithms, the ease of integration with Flask web applications, and the availability of comprehensive documentation and robust developer support. These APIs are renowned for their rigorous testing and validation in the domain of image analysis and search, thus offering dependable and scalable solutions for image processing.

From the myriad APIs suitable for image search, such as Google Cloud, Microsoft Azure, and Amazon AWS, AWS Rekognition \cite{mishra2019machine} emerged as the preferred choice. AWS Rekognition is a cloud-based solution renowned for its sophisticated image analysis capabilities. It leverages deep learning algorithms to undertake visual content analysis, with these algorithms being honed using extensive image databases. This ensures precise identification of patterns and nuances within images. Notably, AWS Rekognition's scalability enables it to execute real-time analysis on extensive collections of images and videos.

Incorporating the ChatGPT API into a search engine can amplify its efficiency. The ChatGPT API can transform user-inputted textual descriptions into structured search queries with its advanced NLP algorithms. This not only refines the search results but also simplifies the user journey in pinpointing the desired images.

In this paper, the ARASAAC\cite{paolieri2018norms} dataset is used to present the message in the image-based and mixed search system. The ARASAAC dataset serves as an invaluable resource in the realm of augmentative and alternative communication, facilitating the conveyance of ideas through a vast array of over 40,000 pictograms and symbols. Sponsored by the Aragonese government, ARASAAC offers symbols that span a broad spectrum of topics, from daily activities and fauna to emotions, sustenance, and transportation modalities. Designed for universal comprehensibility, these pictograms are especially beneficial for individuals with communication challenges. The dataset is readily accessible in various formats, including PNG, SVG, and EPS, through the ARASAAC website or its API. Its significance in the field is underscored by its integration into numerous AAC tools, encompassing speech generators, symbol libraries, and communication boards.

The frontend comprises three segments: the text-based conversational search system, the image-centric conversational search system, and a hybrid system. The text-based system facilitates user interaction via text and voice, whereas the image-centric system focuses on image-based interactions. The hybrid system accepts image inputs and responds with a combination of images and texts\cite{zhou2020design}.

Regarding the backend, the system integrates OpenAI services to produce text-based chat conversations. Concurrently, a separate backend component is responsible for generating image-text chat interactions and offers an image search feature powered by the Google API.

The user interface is pivotal for effective user-system engagement. A carefully designed UI can significantly elevate user experience by creating an intuitive, user-centric space. To tailor the design, thorough interviews and surveys determined user needs, leading to a blend of aesthetic and functionality. While initial designs mirrored ChatGPT's layout, feedback favored a conventional chat interface, with system messages on the left and user messages on the right, ensuring familiarity and clarity.

Post user feedback, attention gravitated towards the visual design, culminating in the selection of the Material Design paradigm, a design lexicon conceived by Google. Material Design\cite{material}, a cohesive amalgamation of guidelines, elements, and tools, facilitates the crafting of engaging UIs.

\begin{figure}[ht]
  \centering
  \includegraphics[width=\linewidth]{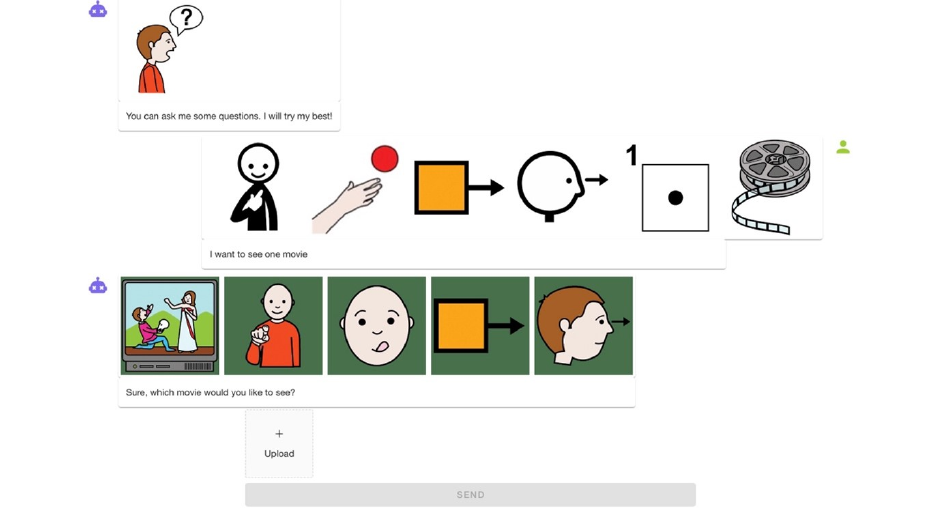}
  \caption{Mixed Search System}
  \label{fig: mix search system}
\end{figure}

The system's architecture adopts a tripartite division, drawing inspiration predominantly from traditional chat application design: system responses are situated on the left, user responses on the right, and user input at the base. Each search functionality, contingent upon its input modality, necessitates a distinct UI design:
\begin{itemize}
    \item Text-based Conversational Search UI: This UI adheres to the archetypal chat application layout with a distinctive feature at its base—an input box coupled with a toggle, enabling users to interchange between text and voice input, facilitated by the web API's speech recognition. 
    \item Image-based Conversational Search UI: Emulating the conventional chat layout, this UI pivots entirely around image responses, incorporating an image upload module at its base for user input.
    \item Mixed Image and Text Conversational Search UI: While this UI's foundation mirrors the image-based counterpart, it integrates text-based responses beneath the images in the system and user sections. Refer to Figure \ref{fig: mix search system} for a detailed visual.
\end{itemize}

Image recognition accuracy was paramount. Initially, labels from images were transformed into sentences using the OpenAI API, but this resulted in unclear sentences. Leveraging the ARASAAC symbol library, where each symbol has a defined meaning, a new method matched input images to library symbols to generate clearer sentences. This refined technique improved the system's conversational fluency.

Adapting the user interface for a diverse user base across various devices was critical. Emphasizing a consistent interface, a responsive design strategy was employed to cater to different screen sizes. Additionally, user feedback mechanisms were integrated, enabling continuous UI/UX improvements.

In summation, sculpting the UI for a conversational search system proved to be an intricate endeavor, but one that was immensely gratifying. The resultant UI, a product of meticulous attention to user feedback and iterative design, seamlessly marries aesthetics with user-centric functionality.

\subsection{Experiment Design}

Twenty-one volunteers were enlisted for the study. The predominant group comprised 18 students from the University of Southampton, aged between 23 and 25 years. Additionally, there was one female participant under 18 and two participants over 50, one male and one female. None of the participants had visual impairments or neurological or psychiatric disorders. Due to sensor inconsistencies leading to considerable data variation, data from one participant was excluded. Consequently, data from 20 participants was utilized for the study.

The primary objective was to garner sensor data from participants interacting with a search engine across three modalities: Text, Image, and Mixed mode. Preceding the experiment, the seating position relative to the monitor was standardized, accounting for fixed placements of both the camera (resolution: 1920×1080 pixels, 30fps) and the eye-tracker\cite{eyetracker}. This ensured optimal facial capture and eye-tracking data accuracy.

Upon arrival, participants were directed to don the Shimmer sensor, connecting two electrodes to two fingers and an additional electrode to the ear. Once sensors were verified to communicate effectively with the data platform, the experimental tasks commenced. 

A structured flow was followed throughout the experiment, consolidated within a unified interface. This began with an eye-tracker calibration using the iMotions\cite{imotions} software, followed by a brief system introduction on a welcome screen.

Subsequent steps included:
\begin{itemize}
    \item Consent Form: Ensuring ethical compliance by informing participants of research objectives and collecting informed consent.
    \item Demographic Input: Participants entered basic demographic details, including age, gender, education, and any physical conditions.
    \item Instruction Page: A preparatory guide to familiarize participants with the upcoming tasks.
\end{itemize}

The crux of the experiment revolved around participants interacting with a search engine to complete three tasks, which varied in content but maintained a consistent structure. To avoid bias arising from the order of tasks, their sequence was randomized using a card draw system. The tasks are as follows:

\begin{description}
    \item[\textbf{Task 1:}]   Ask the system to recommend a movie and tell you the duration of the movie.
    
    \item[\textbf{Task 2:}] Ask the system to recommend an action movie and tell you the duration of the movie.
    
    \item[\textbf{Task 3:}] Ask the system as follows:
    \begin{itemize}
        \item ``I want to see a movie"
        \item ``Action movie"
        \item ``Tell me the duration of the movie"
    \end{itemize}
\end{description}

Participants were then asked to complete these tasks in a randomized different system of the three systems mentioned above and fill out a questionnaire to reflect their experience with the three search engine modes. The content of the questionnaire is shown in the table


\begin{table}
\caption{User Feedback Survey}\label{tab:User Feedback Survey}
{
\begin{tabular}{|p{1.5cm}|p{7cm}|p{5cm}|}
    \hline
    \textbf{Question} & \textbf{Content} & \textbf{Options} \\
    \hline
    Question1 & Which part of this system would you prefer to use? & A. Text B. Image C. Mixed \\
    \hline
    Question2 & Which part of this system provides more accurate information? & A. Text B. Image C. Mixed \\
    \hline
    Question3 & Which part of this system's UI do you prefer? & A. Text B. Image C. Mixed \\
    \hline
    Question4 & I believe that conversational search only by text and voice could help with intellectual disability & A. Strongly disagree B. Disagree C. Neutral D. Agree E. Strongly agree \\
    \hline
    Question5 & I believe that conversational search only by image could help with intellectual disability & A. Strongly disagree B. Disagree C. Neutral D. Agree E. Strongly agree \\
    \hline
    Question6 & I believe that conversational search by both image and text could help with intellectual disability & A. Strongly disagree B. Disagree C. Neutral D. Agree E. Strongly agree \\
    \hline
\end{tabular}
}
\end{table}

\section{Findings and Discussion}
A collection of 21 data sets was gathered, each encompassing skin conductivity, heart rate, eye movement data, and facial emotion metrics for an individual participant. Upon analysis, one data set displayed substantial deviation, possibly attributable to incorrect sensor placement, and was consequently excluded. This resulted in a compilation of 20 valid data sets, each generating an average of 10 minutes of sensor data. The iMotions software captures sensor data at millisecond intervals, ensuring a precision of 1 millisecond in the obtained CSV files.

The principal objective of this study is to facilitate individuals with physical disabilities, particularly those facing challenges using a mouse, in executing searches via sensors. As such, it is imperative to analyze users' physiological data during their decision-making process in a search operation\cite{white2017improving}. Within our experimental framework, a user's decision is indicated by activating the SEND button in the dialogue system, necessitating an examination of physiological data concurrent with a mouse click\cite{Huang2011NoCN}. To leverage sensor data as a surrogate for mouse actions, it's essential to grasp users' mouse usage patterns. While direct investigation of physiological data at the mouse click instance is feasible, pinpointing the exact moment of action is arduous. Moreover, the sheer volume of data from a singular moment might result in fragmented data, increasing the risk of analytical errors due to potential physiological signal inaccuracies. Consequently, a broader temporal scope is advocated for analyzing user operation behavior. The association between mouse clicks and search rankings, introduced a temporal interval spanning the onset of mouse movement to the subsequent cessation to elucidate user behavioral traits \cite{shi2007galvanic}. Their findings emphasized the optimal duration of 40 milliseconds post-initiation of mouse movement to balance data volume against user behavior insight. Aligning with this, our analysis focuses on sensor data fluctuations within the 40 ms post-mouse movement, utilizing the period's average data as a reflection of users' physiological traits. The subsequent challenge lies in ascertaining the commencement point of this interval.

The study commenced by examining the average GSR (Galvanic Skin Response) levels of users across the three search engine modes. GSR, an indicator of an individual's physiological activation and emotional state, is derived as the mean value over the entirety of the user's interaction period. This data, quantified in microsiemens ($\mu$S) to represent conductivity, is procured from the Shimmer sensor. Notably, conductivity elucidates the resistance encountered by an electric current as it traverses through a unit material area. Within the realm of GSR measurements, this is employed to signify variations in skin conductivity, predominantly attributed to perspiration.

\begin{figure}[ht]
  \centering
  \includegraphics[width=0.7\linewidth]{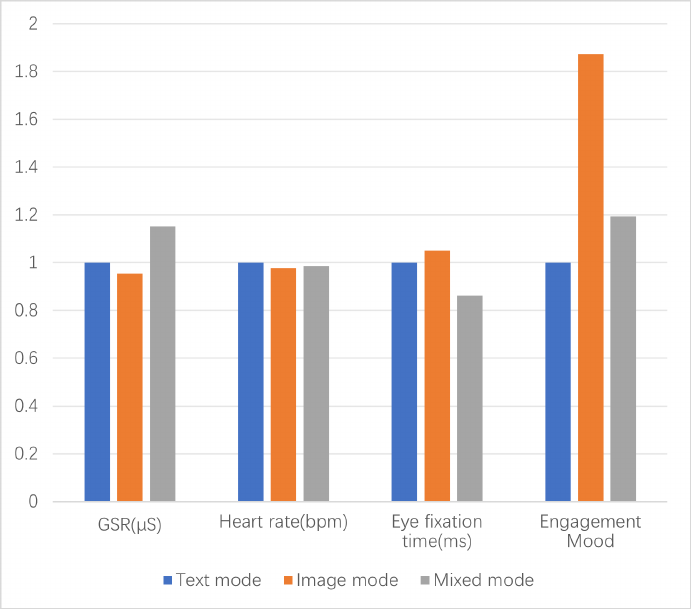}
  \caption{Variation of data in different modes}
  \label{fig: Variation of data in different modes}
\end{figure}

GSR levels escalate in tandem with a user's cognitive engagement when interacting with a computer via voice and gestures \cite{shi2007galvanic}. Elevated GSR levels are typically observed during states of emotional arousal, such as excitement or anxiety, attributed to heightened activity in the sympathetic branch of the autonomic nervous system, resulting in increased sweat production and skin conductance. Conversely, states of relaxation or pleasure may witness a reduction in GSR, due to elevated activity in the parasympathetic branch, leading to diminished skin conductance.


This investigation explored the correlation between the physiological parameters of participants while utilizing various search engine display modes. Through analyzing fluctuations between emotional response and cognitive load, this study aimed to discern if search engines that utilize imagery, or a combination of imagery and text, influence users' cognitive load or emotional states. Figure \ref{fig: Variation of data in different modes} delineates the variance in several datasets across different modes.


To represent the correlation of data across diverse modes, data in Text mode was normalized to 1. Subsequent data was calculated in proportion to this baseline, with results rounded to three decimal places.

It is evident that the form of information (either Text or Image) provided by the search engine doesn't markedly affect heart rate. Conversely, parameters like GSR, eye fixation time, and engagement mood exhibit significant alterations. Specifically, GSR diminishes in Image mode but escalates in Mixed mode, suggesting that a purely image-based display might effectively curtail user stress and cognitive load. However, a combined display appears to augment both factors. From an ocular trajectory perspective, Image mode intensifies comprehension challenges, necessitating users to fixate longer for content assimilation. The Mixed mode, incorporating both visual and textual elements, seems to mitigate this challenge. Regarding facial expressions, both Image and Mixed modes induce notable shifts in users' expressions, potentially because the image format incurs an initial learning curve, leading to evident surprise among first-time users.

\begin{figure}[ht]
  \centering
  \includegraphics[width=0.7\linewidth]{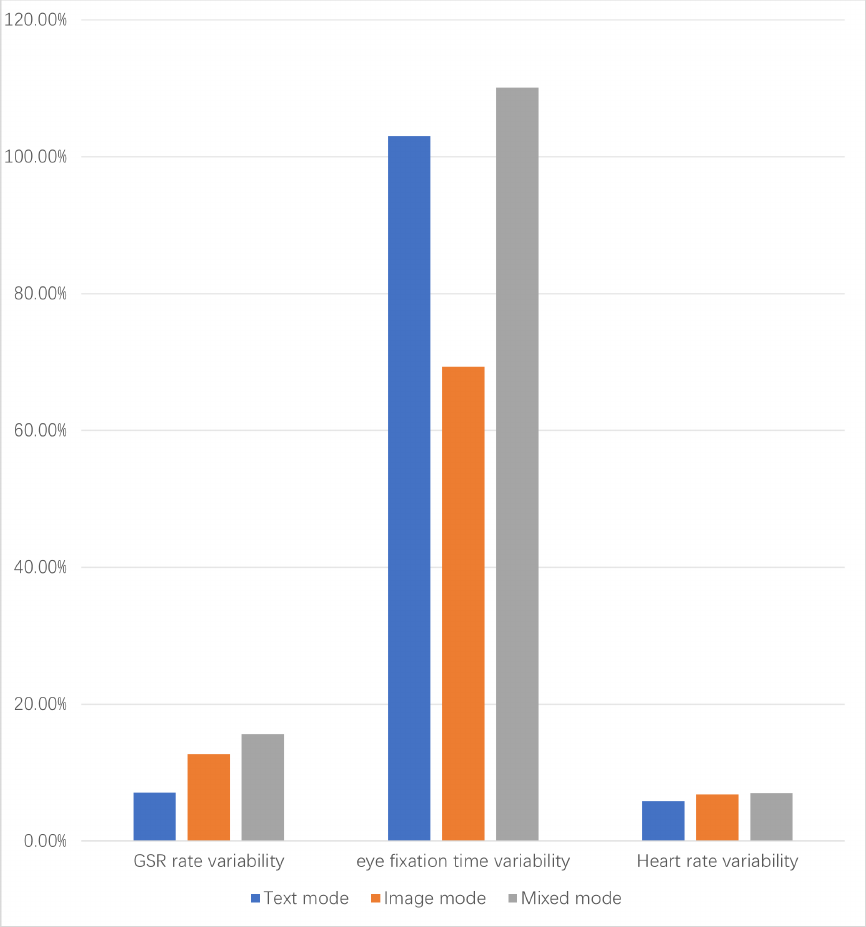}
  \caption{Rate of change in icon form}
  \label{fig: Rate of change in icon form}
\end{figure}

In examining decision-making data shifts, it was noted that within the initial 40 milliseconds of mouse movement, there was a discernible change in physiological signals, presumably indicating a decision-making process. To validate if physiological data can signify decision-making during searches, data procured by sensors in this timeframe were meticulously analyzed. Figure \ref{fig: Rate of change in icon form} highlights the degree of change for each dataset within this interval.


Overall, all three figures rise when users make certain decisions. From the above table, the average growth rate of GSR is 11.8\% in all three modes. And the growth rate of Eye fixation time is 94.15\%. And the average growth rate of Heart rate is 6.53\%.

Upon analyzing the sensor data, several patterns emerged. The physiological signals, especially eye movements and gestures, indicated a high level of user engagement when interacting with the image-based conversational search system. The frequency of positive feedback actions, such as nodding in agreement, was significantly higher compared to instances of confusion or dissatisfaction.


Feedback from participants further verified these findings. A significant number of users expressed a preference for the text-based and mixed systems. Their belief in the potential of conversational search systems, especially in helping individuals with intellectual disabilities, was also verified.

The image-based conversational search system was generally well-received by users. The visual representation of search results provided an intuitive interface, reducing the cognitive load on users. This was particularly evident in the reduced time taken for users to comprehend search results, as compared to traditional text-based outputs. Additionally, the adaptive feedback loop mechanism, which refined search results based on real-time user feedback, further enhanced user satisfaction levels.

One of the standout findings was the system's potential to revolutionize the search experience for users with disabilities. The visual-centric approach catered to those with linguistic or cognitive challenges, making information more accessible. Furthermore, the sensor integration, which allowed for gesture-based feedback, provided an inclusive platform for those with physical disabilities. This was a significant stride towards creating a more equitable digital landscape.

When juxtaposed with traditional text-based conversational search engines, the image-based system showcased several advantages. The most prominent was the system's ability to transcend language barriers, given its reliance on universally understood visual cues. Moreover, the sensor-based feedback mechanism provided a more organic and intuitive user experience, as opposed to the often cumbersome text-based feedback methods. However, it's worth noting that while the image-based system excelled in inclusivity and user engagement, there were instances where users familiar with traditional systems took time to adapt to this novel approach.

In conclusion, the findings underscored the potential of the image-based conversational search system, not just as a novel innovation, but as a tool for fostering inclusivity in the digital realm. The system's design, rooted in user-centric principles, and its adaptive capabilities, driven by sensor feedback, set it apart as a pioneering solution in the realm of information retrieval.

\section{Conclusion and Future Recommendations}

The inception of this project was motivated by the objective of crafting an intuitive user interface, prioritizing ease of use, and fostering a user-centric experience. The trajectory of the project was marked by the adoption of a renowned frontend framework, supplemented by the material design style for frontend development. To enhance the depth of user experience analysis, the iMotions application was harnessed alongside sensors for data collection and assessment.

One of the project's cornerstone achievements was the design and execution of a conversational search system. The resilience of the system's backend was pivotal in ensuring swift data acquisition and processing. Furthermore, the project witnessed significant strides in user interface development, characterized by a natural, dialogue-centric design. Through the implementation of renowned front-end frameworks and the material design aesthetic, users were guaranteed an intuitive interaction paradigm. Optimization of the data structure was instrumental in expediting searches and rendering, assuring users of timely feedback. This interface's versatility across diverse devices underscored its efficacy. 

The system's evaluation phase furnished indispensable insights into both system performance and user experience. Comprehensive assessments were ensured through the deployment of varied task levels, complemented by a counterbalancing strategy. Emotion analysis, underpinned by the iMotions application, illuminated both the system's strengths and areas ripe for enhancement. Participant feedback was overwhelmingly affirmative, with particular emphasis on the system's promise for individuals with intellectual challenges. While each system presents its unique strengths, the hybrid conversational search system seemingly offers an optimal blend of clarity and engagement. The feedback further underscores the transformative potential of these systems in enhancing technological accessibility for individuals with disabilities.

Emotion analysis, conducted using the iMotions application, yielded significant insights into user experience. The text-based search system was observed to be the most intuitive and least perplexing for users. Conversely, the image-based system demanded greater cognitive effort from users, resulting in heightened confusion. Interestingly, the hybrid system, integrating both text and images, received the most engagement, indicating its enhanced interactivity for users.

Participant feedback corroborated these observations. A notable proportion expressed a predilection for both the text-based and hybrid systems. Moreover, the efficacy of conversational search systems, particularly in assisting individuals with intellectual disabilities, received affirmation.

In summary, while each system presents its unique strengths, the mixed conversational search system seemingly offers an optimal blend of clarity and engagement. The feedback further underscores the transformative potential of these systems in enhancing technological accessibility for individuals with disabilities.

This assessment lays a robust groundwork for subsequent enhancements to the system, aiming to cater to a broader audience without compromising its fundamental functionality and user-centric design.

We anticipate the following developments in the next stages of research and improvement: First, our algorithms will be painstakingly fine-tuned to improve search result accuracy and make sure results align more closely with user preferences. Next, we want to expand the scope of our emotion analysis by including a larger number of sensors and a broader range of emotions. In addition, we have observed from feedback that there may be a benefit to enhancing accessibility features, particularly in order to accommodate users with cognitive disabilities, highlighting our dedication to inclusiveness. Finally, we want to strengthen our commitment to always adapting to the changing needs and preferences of our users by adding a real-time feedback mechanism to the system.

\bibliography{sample-1col}




\end{sloppypar}
\end{document}